# Computing Vertex Centrality Measures in Massive Real Networks with a Neural Learning Model

Felipe Grando
Institute of Informatics
Federal University of Rio Grande do Sul
Porto Alegre, Brazil
fgrando@inf.ufrgs.br

Luís C. Lamb
Institute of Informatics
Federal University of Rio Grande do Sul
Porto Alegre, Brazil
lamb@inf.ufrgs.br

*Abstract*—Vertex centrality measures are a multi-purpose analysis tool, commonly used in many application environments to retrieve information and unveil knowledge from the graphs and network structural properties. However, the algorithms of such metrics are expensive in terms of computational resources when running real-time applications or massive real world networks. Thus, approximation techniques have been developed and used to compute the measures in such scenarios. In this paper, we demonstrate and analyze the use of neural network learning algorithms to tackle such task and compare their performance in terms of solution quality and computation time with other techniques from the literature. Our work offers several contributions. We highlight both the pros and cons of approximating centralities though neural learning. By empirical means and statistics, we then show that the regression model generated with a feedforward neural networks trained by the Levenberg-Marquardt algorithm is not only the best option considering computational resources, but also achieves the best solution quality for relevant applications and large-scale networks.

*Keywords—Vertex Centrality Measures, Neural Networks, Complex Network Models, Machine Learning, Regression Model*

## I. INTRODUCTION

The increasing development and ubiquity of large-scale networks poses several technological challenges for both researchers and professionals. Network digital and physical connections such as digital links between webpages, friend lists on social networks, wireless vehicular networks, cable connection between routers, streets paths and energy grids raise a growing number of complex questions. As a result, analyzing and understanding network properties is fundamental in several application domains [1] [2]. Therefore, the development of network analysis tools that provide information about the large number of existing networks is clearly relevant [3] [4].

One particular class of metrics known as vertex centrality are particular relevant, given their general applicability. Such metrics and associated algorithms aim at evaluating, ranking and identifying the importance of vertices (elements of a given network) using only the network basic structural properties as input information [5].

Centrality measures have been used in many areas over the last years, including: computer networks [6] [7], complex networks construction [8] [9], artificial intelligence applications [10] [11], social network analysis [12] [13], traffic and transport flow [14] [16], game theory, problem-solving [17][18][19] and biology [20] [21].

However, some of these metrics are computationally expensive even when their algorithms are polynomial in time; this is due to their use in massive networks with millions or even billions of vertices [22]. To tackle such complexity, we apply and experiment with a methodology that uses a complex network model and machine learning techniques to approximate two of the mostly used centrality measures (namely, *betweenness* and *closeness* centralities) in real world networks. Further, we use as input the information provided by two other centrality measures (*degree* and *eigenvector*).

In order to obtain training samples for the neural learning step, we use the Block Two-Level Erdős and Rényi – BTER [23] [24] generative network model. By using the BTER model, one can generate networks with diminished size, but with the same structural properties of massive real networks. The data obtained with the BTER synthetic networks is proven effective in training the neural network model, which is then able to generalize for the real datasets.

In the coming sections, we also identify the appropriate combination of meta-parameters for an optimized learning process, including insights about the best options for neural network structure, training algorithm, multitasking and training information.

In this paper, we illustrate the effectiveness of the proposed method on thirty real world scenarios comprising networks with sizes ranging from thousands to hundreds of thousand vertices. Finally, we compare the results of the neural learning model with other approximation techniques and with the exact computation of the centrality values. We show how the machine learning methodology is considerably faster than the others are and - at the same time – we show that it renders competitive results in comparable quality with other approximations methods, but not in all experiments.

The remainder of our paper is organized as follows. First, we introduce centrality measures, complex network models and related work in Section II. In Section III we explain our methodology and experimental analysis. Section IV concludes and points out future research avenues.

## II. BACKGROUND AND RELATED WORK

### A. Centrality Measures: A Summary

In a nutshell, centrality measures quantify the centrality of each vertex of a network, usually creating a rank of vertices, or the entire network itself, by computing how centralized a given network is [3] [5]. The term centrality in this context can admit several meanings depending on the application domain under analysis. It may stand for power, influence, control, visibility, independency and/or contribution for instance [1] [3] [4] [5].

The vertex centrality measures are important to identify elements behavior and roles within a network and are usually employed to create ranks among vertices that are used as comparison factors with others domain specific metrics [3] [4]. Such metrics are characterized by deterministic algorithms. Each algorithm tries to capture a distinct *idea* or *concept* of "centrality of a vertex". Each algorithm and measure are distinct from the others although they frequently share a common goal and produce similar/close results. Some of these metrics are adaptations from the others to enable their use on digraphs or weighted networks [3]. In this work, we will use four centrality measures, namely: *degree*, *eigenvector*, *betweenness* and *closeness*. They are the most widely applied metrics and are some of the only metrics in which exact computation it is still possible for massive networks due to their lower complexity. This is important to allow our empirical and statistical analysis on massive networks within feasible time with resources available for our research (the experiments and further justifications are explained in detail in Section III). The centrality $C$ of a vertex $w$ can be computed with the formulas depicted in Table I.

Table I also shows the algorithm time complexity considering a graph with $n$ vertices and $m$ edges. It is important to notice that although the complexity of *eigenvector* is quadratic there is a simpler computation method known as the *power method*, which is an iteration algorithm that approximates the exact metric within few steps and avoids numerical accuracy issues [30].

TABLE I. VERTEX CENTRALITY MEASURES FORMULAE

| Centrality Measure | Formulae | Algorithm Complexity |
|---|---|---|
| Degree [5] | $C_D(w) = \sum_{i=1}^{n} a(i,w)$ | $\Theta(m)$ |
| Betweenness [25] [26] | $C_B(w) = \sum_{i=1}^{n} \sum_{j=i+1}^{n} \frac{g_{ij}(w)}{g_{ij}}$ | $O(mn)$ |
| Closeness [27] [28] | $C_C(w) = \frac{1}{\sum_{i=1}^{n} d(i,w)}$ | $O(mn)$ |
| Eigenvector [29] [30] | $C_E(w) = \sum_{it=1}^{+\infty} \frac{(E^{it} A)_w}{\sum_{i=1}^{n} E_i^{it-1}}$ | $O(n^2)$ |

$a(i,w)$ admits value 1 when vertices $i$ and $w$ are adjacent; $g_{ij}(w)$ is the number of shortest paths between vertices $i$ and $j$ that passes through $w$; $d(i,w)$ is the length (distance) of the shortest path between vertices $i$ and $w$; $A$ is the adjacency matrix of a graph, $E$ is the eigenvector (initialized with ones on iteration 1) and *it* is the current iteration.

### B. On Complex Network Models

In complex networks research, one aims at understanding and proposing models and theories about the formation of real world networks and their structural properties. The field attracted a lot of attention recently, due to the high availability of large networks and connected knowledge bases. Such studies have aimed at understanding networks' common characteristics and at creating models capable of generating networks stochastically with similar properties. These models became a valuable research tool for many disciplines [1] [22].

Several complex network models were developed and studied over the last years. However, some of them are unable to capture and represent some properties of real networks, while others are only conceptualizations and unfortunately are not designed for generating synthetic networks [4] [22]. The Block Two-Level Erdős and Rényi (BTER) model [23] generates networks with very similar properties of real networks. It builds a network based on a desired degree distribution and clustering coefficients (average clustering for each set of vertices with same degree or the global clustering of the network).

The BTER model is divided into three steps [24]:

(i) the vertices are grouped by degree in communities with size equal to the degree of its members plus one;

(ii) each community is considered an Erdős and Rényi graph [31] with a probability of connection among vertices equals to the clustering coefficient;

(iii) connections among communities are generated proportionally to the excess degree of each vertex (number of connections that a vertex needs to match its desired degree). This weighted distribution is similar to the Chung and Lu graphs [32] [33].

### C. On Approximating Centrality Measures

Typical centrality measures algorithms do not scale up to massive graphs (such as large social networks and web graphs). This can be even more critical when one needs to compute many centrality queries, particularly when one is interested in the centrality of all vertices or whenever the network structure is dynamic in time.

Several authors address the high computational complexity of the centrality measures algorithms [34] [36]. They usually propose approximation techniques for specific network metrics based on sampling methodologies in which, the exact computation of the metric is realized for sampled vertices and then is used as reference to estimate the centrality values of the others vertices of the network. For instance, the betweenness and closeness centralities share a quite similar foundation. One computes the *single source shortest path* (SSSP) for a given number of sample vertices. Each SSSP tree gives the exact centrality value for its source vertex. At the same time, one can use them as an approximation for the other vertices considering that all previously computed SSSP trees are partial solutions for such vertices. Therefore, a given vertex not sampled will have its centrality value approximated by an average result given by all SSSP trees from the sampled vertices. An algorithm for such objectives was defined and tested in real case scenarios by Bader et al. [34] for betweenness and by Eppstein and Wang [35] for closeness centralities. However, the simple approach given by the sampling technique has a few drawbacks and leads to relevant questioning such as: how many vertices should be sampled and how should one select them? Or, how can one efficiently parallelize the sampling technique considering that it is no longer possible to compute each vertex centrality independently?

Brandes and Pich [36] studied the vertices selection problem. They proposed many heuristics to choose vertices for the sampling techniques, starting with simple ones, such as picking the high degree vertices and finishing with more complex ones, which consider vertices distances and mixed strategies. Despite their attempts to find an optimal heuristic for such problem, they concluded that picking vertices uniformly at random on average is the best strategy when considering different types of network structures. Some authors [22] [37] [38] tried other kind of approximation methodologies, based mainly on machine learning and neural networks. The strength of these methodologies is their adaptability (they can be used to approximate several distinct centrality measures) and they are able to compute the centrality measures a lot faster than any other method after the model is trained.

The main problems with a neural machine learning approach are their configuration and training for this task. Complex network models were applied to generate synthetic networks providing plentiful training data for the training algorithm. However, one asks how and what is the best option to generate such networks? And, which attributes should be used to train the models? These are still open issues. In order to answer such questions, we put the ideas into a comprehensive experimentation. This will serve to illustrate the capabilities of the method in validation and comparative analyses.

### III. METHODOLOGY AND EMPIRICAL ANALYSIS

The methodology applied in our experiments is divided in four main steps:

(A) acquiring test (real world networks) and training data (synthetic networks generated with the BTER complex network model);

(B) testing and validating the meta-parameters of the artificial neural networks and training algorithms;

(C) computing the centrality measures using the exact algorithms, the model generated by the artificial neural network and the sampling-based algorithms in real networks;

(D) comparing the results of the approximation techniques.

#### A. Acquisition of Testing and Training Data

The testing data was selected from four freely available data repositories: *Stanford large network dataset collection,* *social computing data repository, BGU social networks security research group* and *the Klobenz network collection.*

The selected networks are symmetric and binary (unweighted edges). Only the largest connected component (LCC) was used for the tutorial experiments, which was computed for all analyzed networks. Thirty real networks were selected with sizes ranging from a thousand to 1.5 million vertices.

The BTER complex network model was used to obtain enough and consistent training data. The BTER complex network model was chosen for such a task as it is one of the best models to reproduce real world networks structural properties. Moreover, it is easy to implement and configure and capable to generate plentiful networks with reduced size keeping the most relevant structural properties presented by massive networks.

BTER requires two configuration parameters: the desired degree distribution and the clustering coefficient (which can be configured as a global value, by vertices degree or by community structure).

In our experiments, we applied both a heavy-tailed and a lognormal distribution (parameters provided by [37] [44] [45]) as degree distribution to provide generic data to train and learn the model. Both distributions are known as the best representatives of most real networks degree distribution studied in the literature [44] [45]. For these networks the clustering coefficients were chosen at random in the interval [0,0.7]. A total of 300 generic networks with sizes ranging from a hundred to a thousand vertices were generated at this step. We also generated networks with the degree distributions of six real networks (described in Table II) with their respective clustering coefficients to test the effect of specific vs. generic training data in the machine learning model. These networks were selected as they presented the worst (Amazon, Euroroad, Facebook and PowerGrid networks) and the best (Blog3 and Foursquare networks) overall results obtained by the model trained with the generic training data amongst the thirty real networks selected.

A transformation function (parameters estimated for each degree distribution using as reference a heavy-tailed function model) was applied in the degree distribution to reduce the size of the generated networks. A total of 2400 specific networks (400 for each set of parameters), half with 2000 vertices and half with 3000 vertices.

TABLE II. REAL NETWORKS DESCRIPTION

| Network | Type | Vertices | Edges |
|---|---|---|---|
| Amazon [39] | Co-Purchasing | 334,863 | 925,872 |
| Facebook [12] | Social | 4,039 | 82,143 |
| Blog Catalog 3 [40] | Social | 10,312 | 333,983 |
| Foursquare [40] | Social | 639,014 | 3,214,986 |
| US Power Grid [41] [42] | Supply Lines | 4,941 | 6,594 |
| Euroroad [42][43] | Road | 1,174 | 1,305 |

## B. Parameters Configuration and Validation

We applied the neural network toolbox from Matlab 2015 to configure and train the neural networks. This toolbox comprises most of the algorithms that are well established in the field with a built-in parallelism pool and a robust implementation for industrial, academic and research use.

Following the results in [37] and [38] we selected a fully-connected multi-layer perceptron artificial neural network trained by the Levenberg-Marquardt algorithm (LM) [47].

We tested a full set of combination of parameters with 10-fold cross-validation using the generic training set. The parameters tested were: number of neuron in each layer (from 2 to 100), number of hidden layers (from 1 to 5), the Marquardt adjustment (*mu* - from $5.10^{-6}$ to $5.10^{-3}$), *mu* decrease factor (from 0.1 to 1) and *mu* increase factor (from 1 to 25).

Additionally, we set the activation function of all hidden layers to hyperbolic tangent (any sigmoid function would suffice for such task) and the activation function of the output layer as a linear function (due to the regression task in hand). The batch method was used to update the training parameters.

Our objective here was to find out the most efficient configuration of parameters considering both its computational costs and solution quality. The quality of the solution was measured by the determination coefficient ($R^2$) using only the test set results, which considers 10% of total available training data.

In addition, we also computed the Kendal τ-b correlation coefficient in order to evaluate the quality of the solution of the networks. The Kendall's correlation is a nonparametric measure of strength and association (interval [-1,1]) that exists between two variables measured on at least an ordinal scale. Notice that for most applications the ordering (rank) of the vertices by their centrality values is more important than the centrality value itself.

Each centrality measure uses complex properties of the graph representing the network and the computation of most of these properties are the main reason that centrality measures are time expensive. For such reason, we selected the fastest centrality measures (degree and eigenvector), which are computationally feasible even for massive networks and are highly related to the other metrics [3]. Therefore, we computed both centralities (degree and eigenvector) and ranked the vertices. Such information is then used as input data to train the neural network. We select the vertices rank produced by betweenness and closeness centralities as desired values.

In the experiments, a three-hidden layer network with 20 neurons in each layer achieved the best performance results (considering quality and training speed). Larger networks achieved equally good results in quality but at the expanse of more computational resources. Excluding the *mu* decrease factor that is optimal when configured inside the interval [0.1,0.5], the other parameters of the LM algorithm showed little influence on the overall results displaying no statistical relevance.

This configuration was used for the final training of the artificial neural network. The final training used the same training data (the generic synthetic networks) but with 15% as validation set to prevent overfitting (the training algorithm stops when the performance does not improve in the validation set for ten consecutive batches).

## C. Computation of the Approximation Techniques

In the sequel, we computed each of the four exact centrality measures (eigenvector – $C_E$, betweenness – $C_B$, closeness – $C_C$, and degree – $C_D$) for the thirty selected real networks. The computation of eigenvector and degree centralities is sequential, while betweenness and closeness centralities are computed simultaneously with a merged algorithm keeping the same time complexity upper bounds and using parallelism [28]. All algorithms were programmed in C and the parallel computations used the native OpenMP (Open Multi-Processing interface).

The computation of the metrics used a SGI Altix Blade with 2 AMD Opteron 12-core 2.3GHz, 64GB DDR3 1333MHz RAM memory, Red Hat Enterprise Linux Server 5.4. The computation time of degree centrality required less than 1s to compute for all networks. Eigenvector centrality took at most 2min for the largest network, while betweenness and closeness centralities, even using 24 cores in a parallel environment, took 9 days for the largest network (Hyves). We adopted two sample sizes for each network: 2.5% and 5% of the number of total vertices for the computation of the sampling algorithms. The samples were uniformly randomized and five independent trials with distinct random seeds were executed for each sample size for each network.

The computation of the approximation algorithm was run in the same machine, but in a sequential environment because it would require a larger number of dependent variables between different threads. However, the computation of both algorithms shared similar parts; therefore, their computation was simultaneous for performance improvement. The largest network required almost 2 days for their computation with 5% sampled vertices and took about half for the 2.5% sample.

Next, we computed the centrality measures for all thirty real networks with the trained neural network model (Section III B) and used the Kendall correlation as quality parameter. The results showed a great majority of variability: the model performed poorly for some of the networks (Euroroad, Power Grid, Facebook and Amazon for instance) with coefficients lower than 0.4 and really well in others (Blog3 and Foursquare) with coefficients above 0.7. We think that this behavior is mainly caused by the data used during the training, which was generic for all networks and unable to fulfill specificities for some of the real networks. For such reason, we trained a specific neural network model for each of the six networks that presented the lower and higher results (Euroroad, Power Grid, Foursquare, Amazon, Blog3 and Facebook) with specific generated networks with the parameters provided by each of these networks (Section III A).

We also tested multitasking neural networks capable of learning both centralities (closeness and betweenness) at the same time. Moreover, we tried to add another attribute for the training. In addition to the degree and eigenvector, we added a metric composed by the sum of the vertex degree with the degree of his neighbors (second level degree).

The experiments comprised all the combinations of sizes of training set networks (2000 or 3000), three or just two attributes (addition second level degree or not) and the use of multitasking or not. We generated networks with the same size of the original network exclusively for the Euroroad real network due to its original size of less than 2000 vertices.

To simplify the visualization of the comparative analysis we used the code NN (baseline) for the neural network model trained with the generic dataset and the code NNTAM for the model trained with the specific training set. T assumes value 2 for the networks with size 2000 and 3 for the size 3000, A assumes the quantity of attributes used (2 or 3) and M assumes 1 for simple tasking and 2 for multitasking networks.

### D. Comparative Analisis

First, we have compared and analyzed the correlation values between the approximation methods. The sampling techniques performed better than all neural models tested but the difference is minimal is some cases. The neural models trained with generic networks performed considerably worse than the ones trained with the specific training set of networks excluding one case (Facebook network), where the generic model performed a little better. The difference among the neural models is greater on networks where the first model performed worse.

Amongst the parameters tested, we noticed that the size of the specific networks used for training the model was statistically irrelevant for the results. This was also true in most cases for the multitasking. The addition of a third attribute seems to contribute for the overall performance when approximating the closeness centrality, but sometimes it is harmful to approximate betweenness.

Tables III and IV summarizes the correlation results of some of the combinations tested with best neural network model for each network highlighted in gray. The other tests (3xx and 231) were omitted because they generated results similar to their counterparts.

TABLE III. CORRELATIONS COEFFICIENTS FOR BETWEENNESS

| Network | Approximation Technique | | | | | | | |
|---|---|---|---|---|---|---|---|---|
| | Sample | | Neural Network Model | | | | | |
| | 2.5% | 5.0% | NN | 211 | 212 | 221 | 222 | 232 |
| Amazon | 0.91 | 0.91 | 0.27 | 0.35 | 0.32 | 0.26 | 0.26 | 0.27 |
| Blog3 | 0.89 | 0.90 | 0.73 | 0.80 | 0.80 | 0.71 | 0.70 | 0.71 |
| Euroroad | 0.86 | 0.88 | 0.16 | 0.41 | 0.41 | 0.44 | 0.44 | 0.46 |
| Facebook | 0.67 | 0.75 | 0.36 | 0.30 | 0.35 | 0.24 | 0.20 | 0.29 |
| Foursquare | 0.89 | 0.92 | 0.73 | 0.75 | 0.72 | 0.66 | 0.66 | 0.68 |
| PowerGrid | 0.93 | 0.92 | 0.21 | 0.57 | 0.51 | 0.43 | 0.42 | 0.45 |

TABLE IV. CORRELATION COEFFICIENTS FOR CLOSENESS

| Network | Approximation Technique | | | | | | | |
|---|---|---|---|---|---|---|---|---|
| | Sample | | Neural Network Model | | | | | |
| | 2.5% | 5.0% | NN | 211 | 212 | 221 | 222 | 232 |
| Amazon | 0.99 | 0.99 | 0.10 | 0.42 | 0.52 | 0.66 | 0.65 | 0.65 |
| Blog3 | 0.95 | 0.96 | 0.69 | 0.89 | 0.89 | 0.92 | 0.93 | 0.93 |
| Euroroad | 0.87 | 0.90 | 0.09 | 0.60 | 0.60 | 0.62 | 0.62 | 0.62 |
| Facebook | 0.87 | 0.93 | 0.26 | 0.38 | 0.34 | 0.35 | 0.36 | 0.43 |
| Foursquare | 0.93 | 0.94 | 0.48 | 0.85 | 0.84 | 0.88 | 0.88 | 0.88 |
| PowerGrid | 0.90 | 0.93 | 0.12 | 0.03 | 0.14 | 0.26 | 0.26 | 0.24 |

Table V compares the mean determination coefficient ($R^2$) of the NN baseline model with models generated with the same training data but with other machine learning algorithms using 10-fold cross-validation. Notice that even small differences in the $R^2$ are considerable due to large amount of data. We also see that the NN model performed consistently better than all other methods considering that the 99% confidence intervals for all algorithms lie on the fourth decimal place.

TABLE V. REAL NETWORKS DESCRIPTION

| Algorithm | $R^2$ | |
|---|---|---|
| | Closeness | Betweenness |
| NN | 0.97 | 0.92 |
| Linear Regression | 0.95 | 0.87 |
| Coarse Tree | 0.95 | 0.89 |
| Gaussian SVM | 0.96 | 0.88 |

Next, we computed and analyzed the percentage of correctly classified vertices by their rank considering percentile sets of the network. For this analysis, we ordered the vertices by their exact centrality values, divided such set of vertices in percentiles (from 0.2% of the first ranked vertices to the first 25% vertices), and computed the percentage of these vertices for each percentile that appears in the same percentile in each one of the approximations techniques ranked the vertices. A 100% match means that every element from one set is on the other (perfect classification), while 0% means that both sets are completely disjoint.

Such kind of analysis is important to give us an idea of how close/distant the rankings of the vertices are considering only the more important vertices of the network (generally speaking, the ones of interest for most applications). It also shows us better and gives insights "where" the mistakes are, fact that is obscured when considering only the correlation coefficients.

Figure 1 compares the results of the approximation methodologies for each network. Only the best approximation techniques (5% sample and best neural models) are showed to facilitate the reading of the figures. We already expected that the sampling-based techniques performs better than the machine learning models simply because they have access to more information about the overall network structure with the drawback of requiring a lot more of computation time to acquire such information.

We can see that the Blog3 network was the only case scenario where the neural models performed as good as or a little better that the sampling methodologies. The Blog3 network probably has a simpler structure, therefore the information contained in the Degree and Eigenvector centralities (used as training inputs) fully characterize the network structure while in the other networks more complex information is needed.

One should also consider that the neural learning model is capable to compute the centrality for all vertices of a given network in seconds (even for massive networks) and that the sampling techniques takes at least some minutes for the smaller networks and hours or even days for the biggest networks.

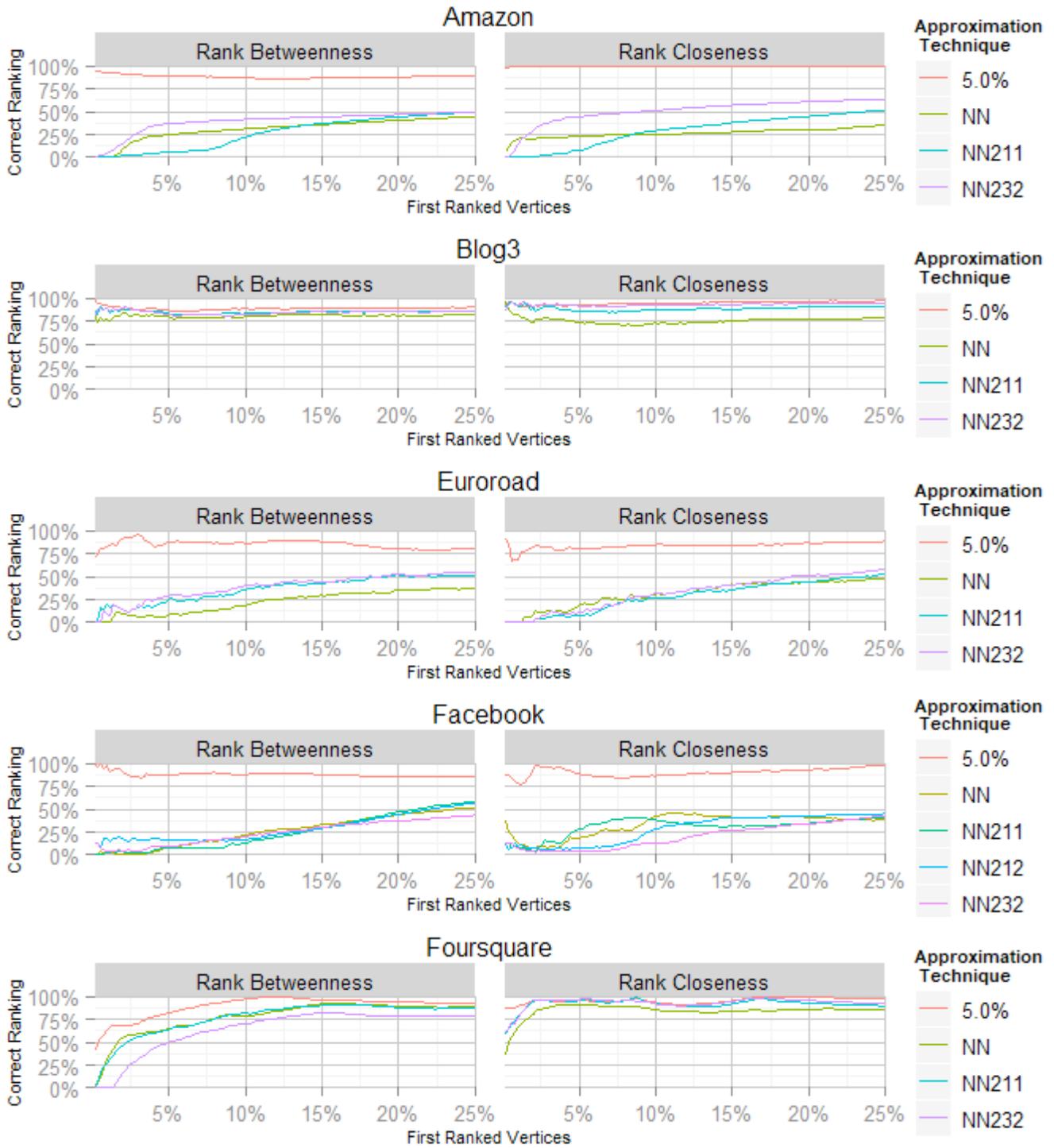

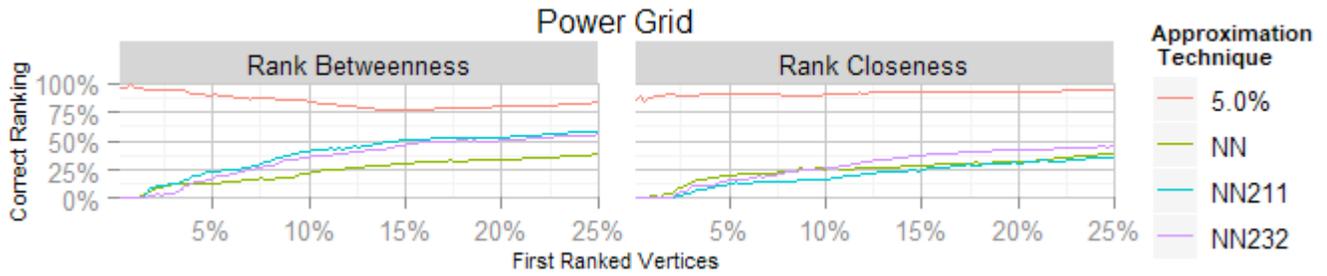

Fig. 1. Percentage of correct classified vertices by percentil set of the network

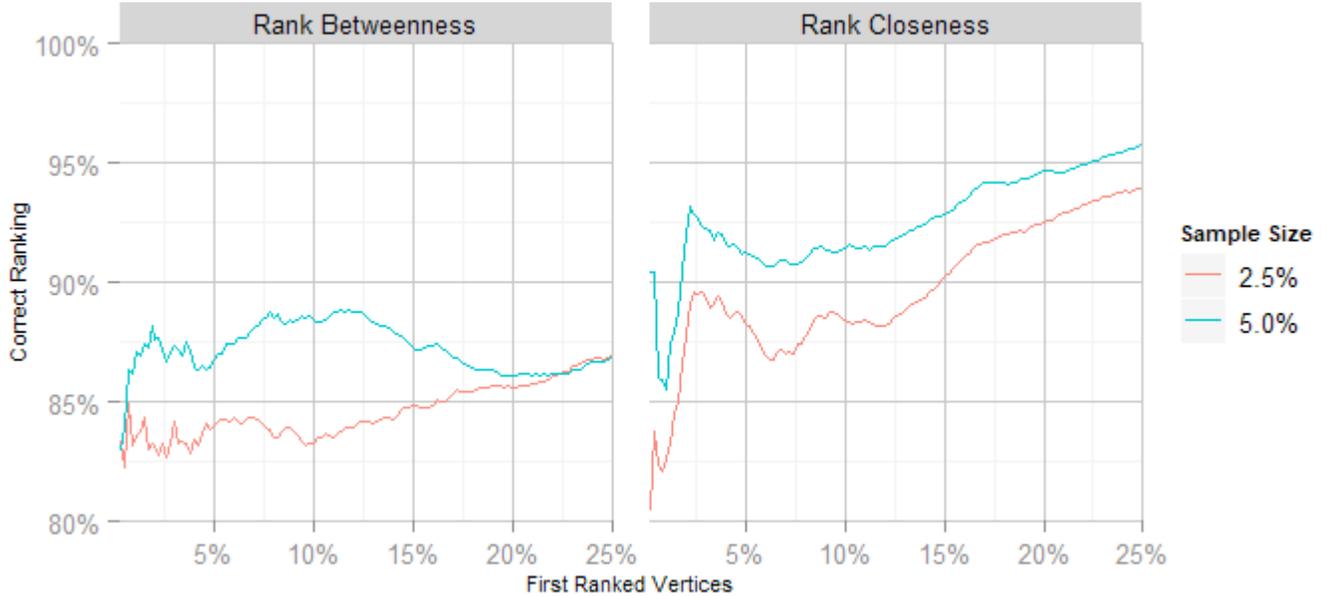

Fig. 2. Comparison between different sampling sizes

In addition, the NN232 configuration seems better than the other NN architectures to rank the first 25% ranked vertices except for betweenness centrality in Foursquare network and for both centralities in Facebook network.

Figure 2 presents the effect of different sampling sizes considering the mean results over the six networks analyzed. It only reinforces what the correlation values already show (Tables III and IV). The upgrade in solution quality for the 5% sample is minimal comparing that it costs twice in terms of computational time.

## IV. Conclusions

The growing relevance of network research and applications demand the development of appropriate tools and methods for network analysis. These methods include vertex centrality measures. However, as networks grow in size, their computational costs present challenges which may hinder some important applications. Machine learning techniques have recently been successful in a number of relevant applications tackling large amounts of data [46] [47] [48]. Moreover, the referred growing availability of massive network databases demands the use of effective techniques to better exploit and to interpret these data. In this context, we presented a comprehensive empirical analysis on the use of artificial neural networks learning to estimate centrality measures. We have tested and identified the best configuration for the artificial neural network training, including network structure, training algorithm, and training meta-parameters. The experimental results revealed that the neural network model is able to approximate the target centrality measures with considerable accuracy and reduced computational costs in 30 real-world experimental case scenarios.

Our research also shows that the data used for training the model is one major factor that affects the learning model. In the real world experiments where generated networks with specific parameters were applied during the training, the performance of the neural model improved considerably. Considering this result, one should always use the knowledge (degree distribution and clustering coefficients) about the network of interest to generate specialized training data; this will lead to improvements in the performance of the artificial neural learning model.

The model showed a noticeable and clear advantage and tradeoff with respect to computational costs, making it a viable option for applications where accuracy is not the only fundamental goal, but in scenarios and configurations in which computation resources are limited. In such common situations, approximations via machine learning are an effective alternative, in particular in the context of large scale complex network applications.